# 4D scale-dependent Schwarzschild-AdS/dS black holes: study of shadow and weak deflection angle and greybody bounding

Ali Övgün[1,a], Reggie C. Pantig[2,b], Ángel Rincón[3,c]

[1] Physics Department, Eastern Mediterranean University, via Mersin 10, 99628 Famagusta, North Cyprus, Turkey
[2] Physics Department, Mapúa University, 658 Muralla St., Intramuros, 1002 Manila, Philippines
[3] Departamento de Física Aplicada, Universidad de Alicante, Campus de San Vicente del Raspeig, 03690 Alicante, Spain

**Abstract** We investigate the shadow, deflection angle, and the greybody bounding of a Schwarzschild-AdS/dS black hole in scale-dependent gravity. We used the EHT data to constrain the parameter $\epsilon$. We have found that within the $1\sigma$ level of uncertainty, $\epsilon M_0$ ranges at $10^{-11} - 10^{-16}$ orders of magnitude for Sgr. A*, and $10^{-11} - 10^{-17}$ for M87*. Using these parameters, we explored how the shadow radius behaves as perceived by a static observer. Using the known parameters for Sgr. A* and M87*, we found different shadow cast behavior near the cosmological horizon even if the same scaling parameters were used. We explored the deflection angle $\hat{\alpha}$ in the weak field limit for both massive and null particles, in addition to the effect of finite distances using the non-asymptotically flat version of the Gauss-Bonnet theorem. We have found that the $\hat{\alpha}$ strongly depends on massive particles. Tests on $\epsilon$'s effect using Sgr. A* is only along points of low impact parameters. For M87*, only the scaled BH in AdS type can be tested at high impact parameters at the expense of very low value for $\hat{\alpha}$. We study the rigorous bound on the greybody factor for scalar field emitted from black holes in the theory and show the bound on the transmission probability. The effects of the scale-dependent gravity on the greybody factors are investigated, and the results indicate that the bound on the greybody factor in this case is less than the bound for the Schwarzschild black hole.

## 1 Introduction

In four-dimensional spacetime, General Relativity (in what follows GR) is the only viable theory (of gravity) able to satisfy: (i) the requirements of diffeomorphism invariance and (ii) the strong equivalence principle [1]. Albeit General Relativity has a solid theoretical and experimental foundation, there are several additional reasons to investigate alternative theories of gravity. Firstly, from a theoretical perspective, two of the most remarkable problems still present in GR are: (i) the presence of singularities [2, 3], and (ii) the impossibility of a renormalization (following standard processes of quantization) [4]. Secondly, from an observational perspective, it is well-known that the discovery of the dark sectors of the Universe becomes evident the necessity of fundamental physics. At this point, it is required the inclusion of new physics to connect both the ultraviolet (UV) and the infrared (IR) sectors. Such a 'New physics' could maintain the laws of gravity and perhaps will be only necessary for the inclusion of new fields (interacting with ordinary matter) that are felt mainly through gravity. In recent years, gravitational wave (GW) astronomy [5–8] has emerged as an indispensable tool for strong-gravity observations. Thus, GWs are representations in which inspiral and merger of binary compact objects, such as black holes and neutron stars, acquire extreme configurations. In the context of an alternative theory of gravity, we should expect GW waveforms to be disturbed [9]. In particular, the changes appear:

(i) At the energy balance of the system,
(ii) Modifying the oscillation frequency of black holes (and neutron stars),
(iii) Deviating the propagation speed of GWs from the speed of light.

Shortly, GW observations can constrain possible deviations from GR, both in the inspiral-merger and in the post-merger regions of the waveform [10, 11]. Consequently, an extensive study of compact objects in modified theories of gravity is needed and expected because experimental progress allows us to rule out some of those theories.

Of particular interest in classical and alternative theories of gravity is the study of black holes. They are usually considered elementary objects. This idea comes from a significant amount of results, usually encoded into the "no-hair theorem" [12]. Moreover, black holes are essential for classical and quantum gravity. On the one hand, they are a generic prediction of Einstein's General Relativity (and other metric theories of gravity) and, on the other hand, the simplest objects in the Universe. Black holes offer a wide variety of features in which classical and quantum effect coexists in a complex form. In this respect, although it has not been detected yet, Hawking's radiation [13, 14] has a particular spot inside the different effects occurring in a black hole. Up to now, we have a few classic and remarkable examples of black holes in four dimensions. The Schwarzschild solution [15], the Reissner-Nordström solution [16, 17], the Kerr solution [18] and, finally, the Kerr-Newman solution [19]. The solutions mentioned above can certainly be connected. Thus, the Kerr black hole solution is the spinning generalization of the Schwarzschild solution, and, on the other hand, the Kerr-Newman solution is the charged spinning generalization of Reissner-Nordström black hole. These four metrics are often referred to as the "black-hole" solutions of general relativity. The aforementioned examples have been significantly studied, obtaining almost all their properties.

[a] e-mail: ali.ovgun@emu.edu.tr
[b] e-mail: rcpantig@mapua.edu.ph
[c] e-mail: angel.rincon@ua.es (corresponding author)

Interestingly, black holes have been investigated in several dimensions. One of the most remarkable examples found is devoted to the 2+1 dimensional case. The BTZ solution [20–22] as well as its modifications (see, for example [23]) offers a perfect arena to study the vast properties a black hole has. Alternatively, the relevance of higher-dimensional black holes becomes evident when we analyze two main ingredients: different rotation dynamics and the appearance of extended black objects. It is well-known that in more than four dimensions, there is the possibility of rotation in several independent rotation planes. The other aspect of rotation that is modified when the number of dimensions increases is the relative competition between the gravitational and centrifugal potentials [24]. The other novel ingredient in $d > 4$ (and usually absent in lower dimensions -at least in vacuum gravity-) is the presence of black objects with extended horizons. The latter means black strings and, in general, black p-branes.

Up to now, black holes in General Relativity have been profoundly analyzed, ranging from 2+1 dimensions to moving the most realistic case, i.e., 3+1 dimensions, up to considering higher-dimensional extensions as 4+1 gravity (or higher dimensional cases). Although successful, GR has specific problems such as the singularity problem or the incompatibility with quantum mechanics. Such issues make it necessary to extend/modify GR. Among other theories, we have several canonical attempts to modify Einstein's General Relativity. Let us mention the well-known Brans-Dicke theory, in which Newton's constant is identified as a scalar field coupled non-minimally with the Ricci scalar [25, 26]. Another alternative is Asymptotically Safe (AS) gravity, a formalism in which the idea is simple: to obtain a consistent and predictive quantum theory of the gravitational field [27], systematically applied to black hole physics [28]. Interestingly, one of the most remarkable ingredients required in any attempt to consistently include quantum features in gravity is the inclusion of running couplings. Thus, coupling constants appearing at the action level are no longer constant. Instead, they are functions that evolve in spacetime. In this respect, we have several alternatives to be used to replace/improve classical gravity compatible with quantum physics. To name a few, we have

(i) The improved formalism [29]
(ii) The variational parameter settings [30]
(iii) Scale-dependent gravity (see details and seminal references in next section), among others. Thus, this paper will consider a black hole solution obtained in the context of scale-dependent gravity in four dimensions with spherical symmetry in the presence of a cosmological constant.

On the other hand, the most visually striking example of the light bending effect is the curious phenomenon of shadow cast and also gravitational lensing of the black hole. The shadow of the Schwarzschild black hole was first studied by Synge in 1966 [31], then Luminet provided a formula for the angular radius of the shadow and visualized it [32]. Recently, the Event Horizon Telescope (EHT) has taken picture of the first image of the supermassive black hole Messier 87 and later Sagittarius A* [33, 34]. This observation reveals a way to test Einstein's theory of general relativity in a strong field regime where the photon ring and shadow have been seen in the image. In the literature there are many studies on shadow of the black hole and also wormholes to understand the nature of the spacetime and their properties in various modified gravity theories [35–65]. On the other hand, gravitational lensing has been known as a impressive tool to address fundamental problems in astrophysics at different scales [66–73]. In 2008, Gibbons and Werner found a new way to derive deflection angle of the black holes in weak field limits by using Gauss-bonnet theorem on the optical metric for the Schwarzschild black hole [74], then Werner extended it to stationary black holes using the Kerr-Randers optical geometry [75]. Since then, this method of Gibbons-Werner has been used in various papers to show the weak deflection angle of many black holes or wormholes in the literature [39–48, 76–92].

Our main goal in this paper is to test if black holes in Einstein's theory of gravity, such as the Schwarzschild black hole, can be distinguished from those in scale-dependent gravity by studying its shadow, lensing, and greybody bounds. To this aim, we will first constraint $\epsilon$ (scale-dependent parameter) using the available data from the EHT on the shadow diameter. Next is to analyze the shadow radius behavior due to such SD parameter along with the cosmological constant. We will also examine the sensitivity of the weak deflection angle in probing $\epsilon$.

The present manuscript is organized as follows: after a short and relatively compact introduction, we review the Schwarzschild-AdS/dS black hole solution. Subsequently, we move to the scale-dependent version of such a black hole in Sect. 2. After that, we compute the corresponding shadow of the 4D scale-dependent Schwarzschild-AdS/dS black hole in Sect. 3. Then, in Sect. 4, the weak deflection of light is investigated in detail. Section 5 is dedicated to calculating the greybody bounding of the new solutions. All results are also shown in figures for better comprehension.

## 2 Scale-dependent Schwarzschild-AdS/dS geometry

In this section, we will briefly summarize the main equations required to describe the physics of two concrete cases:

(i) The classical solution, i.e., the Schwarzschild-AdS/dS black hole solution in the context of General Relativity, and
(ii) The scale-dependent, i.e., the Schwarzschild-AdS/dS solution using the scale-dependent formalism based on asymptotically safe gravity. Both solutions must be mentioned because it is mandatory first to understand which parts of the classical solution are modified to include the corresponding quantum features.

2.1 Classical background

To review the classical black hole solutions in four-dimensional spacetime, we will start by considering the classical Einstein-Hilbert action, $I_0[g_{\mu\nu}]$, with cosmological constant, i.e.,

$$I_0[g_{\mu\nu}] = \int d^4x \sqrt{-g} \left[ \frac{1}{2\kappa_0} \left( R - 2\Lambda_0 \right) + \mathcal{L}_0^M \right], \tag{1}$$

where, as always, $g_{\mu\nu}$ in the metric field, $g = \det(g_{\mu\nu})$ is the determinant of the metric tensor, $R$ is the Ricci scalar, $G_0$ and $\Lambda_0$ represent the classical (i.e., non-scale-dependent) gravitational and cosmological coupling, respectively. In addition, $\kappa_0 \equiv 8\pi G_0$ is the Einstein coupling and $\mathcal{L}_0^M$ is the Lagrangian density for the matter content. For simplicity, we will focus on the simplest case in which the Lagrangian density is taken to be zero. To obtain the classical Einstein's field equations, we take advantage of the action principle with respect to the metric field, namely $\delta I_0/\delta g^{\mu\nu} = 0$, to obtain, for a non-vanishing cosmological constant ($\Lambda_0 \neq 0$), the following equations:

$$R_{\mu\nu} - \frac{1}{2} R g_{\mu\nu} + \Lambda_0 g_{\mu\nu} = 0. \tag{2}$$

Considering a line element of the conventional form

$$ds^2 = -f_0(r)dt^2 + f_0(r)^{-1}dr^2 + r^2 d\Omega^2, \tag{3}$$

and with the help of the Einstein field equations, we can obtain the corresponding lapse function $f_0(r)$, being $r$ the radial coordinate and $d\Omega^2$ is the usual line element of the unit 2-sphere, defined as:

$$d\Omega^2 = d\theta^2 + \sin^2\theta d\phi^2. \tag{4}$$

It should be noticed that the Schwarzschild-AdS/dS solution [93] is a consequence of the effect of the gravitational field on a point-like mass $M_0$. The corresponding classical lapse function is defined as

$$f_0(r) = 1 - \frac{2M_0}{r} - \frac{1}{3}\Lambda_0 r^2. \tag{5}$$

Please, be aware and notice that the sub-index 0 denotes classical quantities. As always, to compute the black hole horizon, we also need to obtain the roots of $f(r = r_0) = 0$, i.e.,

$$r_0^3 - \frac{3r_0}{\Lambda_0} + \frac{6M_0}{\Lambda_0} = 0. \tag{6}$$

The last equation is cubic, and its discriminant, determines the number of reals and imaginary roots could appears. In particular, when its values is taken negative, provides three different real roots. To be more precise, the discriminant $D$ computed by

$$D \equiv R^2 + Q^3, \tag{7}$$

$$R = -\frac{3M_0}{\Lambda_0}, \tag{8}$$

$$Q = -\frac{1}{\Lambda_0}, \tag{9}$$

is equivalent to

$$\Lambda_0 M_0^2 < \frac{1}{9}. \tag{10}$$

Be aware and notice that there are i) a real negative root and ii) two real positive roots corresponding to an event horizon $r_0$, and a cosmological horizon $r_c > r_0$. The particular cases with $\Lambda_0 = 0$ and $M_0 = 0$, the Schwarzschild solution and the maximally symmetric spacetime corresponding to the de-Sitter solution are recovered.

### 2.2 Scale-dependent geometry

We will now move to the scale-dependent version of the above-mentioned black hole. As was previously mentioned in the introduction, to include quantum features into a certain classical background, it is (usually) necessary to allow the classical coupling constant to evolve to functions, which, in principle, run with the energy scale $k$. Up to now, there are a huge variety of well-motivated scenarios where scale-dependent couplings are the key ingredient to obtaining self-consistent solutions. Technically, the starting point is the same that the classical case. However, instead of considering the classical action, we use an average effective action $\Gamma[g_{\mu\nu}, k]$. Taking variations with respect to the metric field, $g_{\mu\nu}$, and the renormalization scale, $k$, we can get the effective Einstein's field equations plus a consistent relation to close the system. Thus, the action in this effective gravity is given as follows:

$$\Gamma[g_{\mu\nu}, k] = \int d^4 x \sqrt{-g} \left[ \frac{1}{2\kappa_k} \left( R - 2\Lambda_k \right) + \mathcal{L}_k^M \right], \tag{11}$$

where the set of parameters, symbolically represented by $\mathcal{O}_k$, have the same meaning as the classical solution although they now depend on the scale $k$. The scale-dependent formalism has been used along the years in several contexts (see [94–127] and references therein). In scale-dependent gravity, the fundamental object, $\Gamma[g_{\mu\nu}, k]$, replaces the classical action $I_0[g_{\mu\nu}]$, and the main difference appears by the inclusion of scale-dependent couplings. Symbolically we can write

$$\{A_0, B_0, (\cdots)_0, Z_0\} \rightarrow \{A_k, B_k, (\cdots)_k, Z_k\}, \tag{12}$$

where the left-hand side of Eq. (12) represents the classical coupling, and the right-hand side collects the scale-dependent couplings. At this point, it is mandatory to mention the physical meaning of the renormalization scale $k$. The connection between $k$ and any physical variable is not well-established, being always a technical and conceptual issue. In curved spacetimes, the concrete form of $k \equiv k(r)$ is not evident [128], different from those occurred in flat spacetimes (where $k \sim r^{-1}$). Naively, we could consider a few ways to make progress regarding $k(r)$. Firstly, we could use semiclassical approaches where a classical metric must be used. The disadvantage, however, is that such a metric is no longer valid in the strong-curvature regime. Another possibility emerge when we consider a $k(r)$ as a function of invariants of the theory, for instance: (i) the Ricci scalar or, (ii) the Kretschmann scalar. Thus, we could introduce a relation as $k \equiv k(R)$, but the concrete form is still missing. These kinds of problems motivated us to review how $k$ is related to variables of the system under study.

Irrespectively of the concrete form of $k$, the scale-dependent formalism contains solutions coming from GR after considering the appropriated limit values. The latter is true by construction, given that SD gravity slightly modifies classical black hole solutions when varying scale-dependent coupling functions replace classical couplings. These kinds of systems (where the coupling constant evolves) can be solved by knowing the concrete form of the coupling constants [94] or, alternatively, from a background independent integration of the functional renormalization group [129–132].

In what follows, we proceed with the final step used in the scale-dependent formalism: we bypass a choice of $k(r)$ recognizing that the coupling functions finally depend on the radial coordinate only (in spherical symmetry):

$$\{G_k, \Lambda_k\} \rightarrow \{G(k(r)), \Lambda(k(r))\}, \tag{13}$$

and then the coupling constants are replaced according to

$$\{G_0, \Lambda_0\} \rightarrow \{G(r), \Lambda(r)\}. \tag{14}$$

The scale-dependent field equations (obtained from a variation of (11) with respect to $g_{\mu\nu}(x)$) are obtained as (see, for instance [97]):

$$R_{\mu\nu} - \frac{1}{2} R g_{\mu\nu} + \Lambda(r) g_{\mu\nu} = -\Delta t_{\mu\nu}. \tag{15}$$

Notice the additional term, $\Delta t_{\mu\nu}$, encodes the G-varying part of the energy-momentum tensor. The concrete form of $\Delta t_{\mu\nu}$ is

$$\Delta t_{\mu\nu} = G(r) \left( g_{\mu\nu} \Box - \nabla_\mu \nabla_\nu \right) G(r)^{-1}. \tag{16}$$

To complete the set of differential equations, a consistency relation can be obtained taking variations respect the field $k$, i.e., $\delta\Gamma/\delta k = 0$. This condition is, however, quite complicated to use to compute exact solutions. Instead, we use the saturated version of the null energy conditions in spherical symmetry (see, for instance, [133]). Specifically, we have [97]

$$2 \frac{G(r)''}{G(r)'} - 4 \frac{G(r)'}{G(r)} = 0, \tag{17}$$

and, as can be observed, this condition does not contain the metric potentials, so, irrespective of the form of $g_{rr}$ and $g_{tt}$, (when $g_{rr} g_{tt} = -1$) the explicit form of Newton's function will always be the same. Solving the differential equation for $G(r)$ we obtain, as previous works [102]

$$G(r) = \frac{G_0}{1 + \epsilon r}, \tag{18}$$

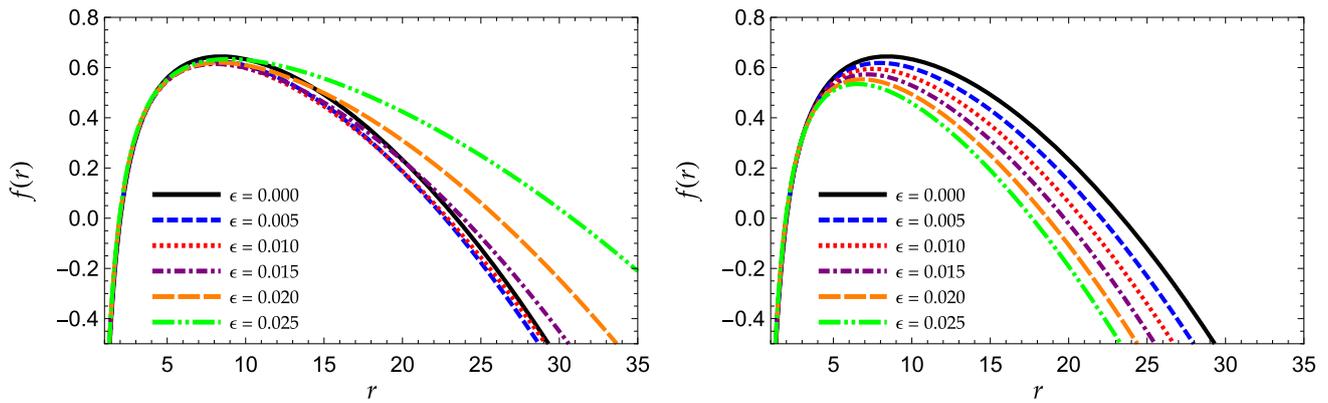

**Fig. 1** Plot of the lapse function $f(r)$ as a function of the radial coordinate. We are considering the following set of values: $G_0 = 1$, $M_0 = 1$ and $\Lambda_0 = +0.005$. The color code is given in figures. **Left Panel:** exact analytical lapse function for the above-mentioned values. **Right Panel:** First order approximation for lapse function for the same values

where $\epsilon$ is the SD parameter and $G_0$ has the classical meaning. Be aware and notice that the classical result is recovered in the limit $\epsilon \to 0$. Now, in the present case, we will consider the line element as follow

$$ds^2 = -f(r)dt^2 + f(r)^{-1}dr^2 + r^2(d\theta^2 + \sin^2(\theta)d\phi^2), \tag{19}$$

and, taking advantage of knowing what $G(r)$ is, we use the effective field equations to get the r-varying cosmological constant $\Lambda(r)$ as well as the lapse function $f(r)$. These expressions are found to be [114]

$$\Lambda(r) = \Lambda_0 + \frac{\epsilon r}{1+\epsilon r}\Lambda_0 + \epsilon^2 \left[\frac{3G_0 M_0}{r}\frac{(1+12\epsilon r(1+\epsilon r))}{(1+\epsilon r)^2}\right.$$
$$+ \frac{2 - 3r\epsilon(6r\epsilon + 5)}{2(r\epsilon + 1)^2} - 3(1 + 6G_0 M_0 \epsilon) \times$$
$$\left. \times \left(\frac{1+2\epsilon r}{1+\epsilon r}\right)\ln\left(1+\frac{1}{r\epsilon}\right)\right], \tag{20}$$

$$f(r) = f_0(r) + \frac{1}{2}\epsilon\left[6G_0 M_0 - 2r + 3r\epsilon(r - 4G_0 M_0)\right.$$
$$\left. + 2r^2\epsilon(1 + 6G_0 M_0 \epsilon)\ln\left(1+\frac{1}{r\epsilon}\right)\right], \tag{21}$$

and taking the limit $\epsilon \to 0$, we obtain

$$\lim_{\epsilon \to 0} \Lambda(r) \to \Lambda_0, \tag{22}$$

$$\lim_{\epsilon \to 0} f(r) \to f_0(r), \tag{23}$$

$$\lim_{\epsilon \to 0} G(r) \to G_0 \equiv 1. \tag{24}$$

By definition, and as a final remark, we can expand the lapse function $f(r)$, at first order in power of $\epsilon$, namely

$$f(r) \approx f_0(r) - \left(1 - \frac{3M_0}{r}\right)(\epsilon r) + \mathcal{O}(\epsilon^2). \tag{25}$$

In Fig. 1, we show the exact lapse function $f(r)$ (left panel) and the first order lapse function (right panel) for the classical solution ($M_0\epsilon = 0$) as well as for five different values of $M_0\epsilon$, namely $M_0\epsilon = 0.005, 0.010, 0.015, 0.020, 0.025$.

The complexity of the lapse function is always a disadvantage because it makes it impossible to obtain analytical expressions for the roots of the algebraic equation $f(r) = 0$. For small $\epsilon$, however, the above expansion may be used to obtain the condition for the existence of three real and unequal roots. Thus, up to the first order in $\epsilon$, we find

$$\Lambda_0 M_0^2 < \frac{1}{9} + \mathcal{O}(\epsilon^2), \tag{26}$$

which means that the classical bound is still valid.

## 3 Shadow cast of 4D scale-dependent Schwarzschild-AdS/dS black holes

Let us now explore the shadow of the 4D scale-dependent Schwarzschild-AdS/dS black hole. The model we used in this study is non-spinning. Thus, we expect the shadow to be a perfect circle if plotted in celestial coordinates. Here, we aim to derive the expression for the shadow radius and analyze its behavior with respect to a static observer at some distance $r_o$ from the black hole. Note that the metric is spherically symmetric, so it would suffice to analyze the equatorial plane (i.e., $\theta = \pi/2$). The Lagrangian for light rays is

$$\mathcal{L} = \frac{1}{2}\left(-f(r)\dot{t} + f(r)^{-1}\dot{r} + r^2\dot{\phi}\right),\tag{27}$$

and implementing the Euler-Lagrange equation, we can find the two important constants of motion as

$$E = f(r)\frac{dt}{d\lambda}, \quad L = r^2\frac{d\phi}{d\lambda}.\tag{28}$$

Immediately, we can now define the impact parameter that simplifies the proceeding calculations further:

$$b \equiv \frac{L}{E} = \frac{r^2}{f(r)}\frac{d\phi}{dt}.\tag{29}$$

With $ds^2 = 0$ for null rays, it is easy to find the orbit equation. That is, an expression that gives information on how the $r$-coordinate changes with $\phi$-coordinate:

$$\left(\frac{dr}{d\phi}\right)^2 = \frac{r^2}{f(r)^{-1}}\left(\frac{h(r)^2}{b^2} - 1\right),\tag{30}$$

where the definition of $h(r)$ is [61]

$$h(r)^2 = \frac{r^2}{f(r)}.\tag{31}$$

Next, the photonsphere radius $r_{ps}$ can be sought off under the condition $h'(r) = 0$. In our case, we find it as

$$r_{ps} = \frac{1 + 3\epsilon M_0 - \sqrt{9\epsilon^2 M_0^2 + 1}}{\epsilon},\tag{32}$$

which is the only solution that gives physical significance.

In Ref. [61], we can find easily that the definition of the angular radius of the shadow when the observer is at $(t_o, r_o, \theta_o = \pi/2, \phi_o)$ is

$$\tan(\alpha_{sh}) = \lim_{\Delta x \to 0}\frac{\Delta y}{\Delta x} = \left(\frac{r^2}{f(r)^{-1}}\right)^{1/2}\frac{d\phi}{dr}\bigg|_{r=r_{obs}},\tag{33}$$

or, for instance, it can be simplified as

$$\sin^2(\alpha_{sh}) = \frac{b_{crit}^2}{h(r_{obs})^2},\tag{34}$$

using the orbit equation in Eq. (47). Indeed, this equation needs the critical impact parameter $b_{crit}$, which in general can be found under the condition that $dr^2/d^2\phi = 0$. For any static and spherically symmetric spacetime, this can be derived using [46]

$$b_{crit}^2 = \frac{4r_{ps}^2}{rf'(r)|_{r=r_{ps}} + 2f(r_{ps})}.\tag{35}$$

We get the result as

$$b_{crit}^2 = \frac{12r_{ps}^3}{r_{ps}(18\epsilon M_0 + 6) - 9\epsilon r_{ps}^2 - 4\Lambda_0 r_{ps}^3 - 6M_0},\tag{36}$$

and we can then derive the shadow radius as

$$R_{sh} = \left\{\frac{12r_{ps}^3}{r_{ps}(18\epsilon M_0 + 6) - 9\epsilon r_{ps}^2 - 4\Lambda_0 r_{ps}^3 - 6M_0}\left[1 - \frac{2M_0}{r_{obs}} - \frac{\Lambda_0 r_{obs}^2}{3} - \left(1 - \frac{3M_0}{r_{obs}}\right)\epsilon r_{obs}\right]\right\}^{1/2}.\tag{37}$$

With this formula, we can now form constraints to $\epsilon$ based on EHT data. Let us now discuss the observational constraint of the parameter $\epsilon$ using the obtained data from M87* [33] and Sgr. A* [34] in relation to black hole shadow. These important observational data are in Table 1.

**Table 1** Black hole observational constraints

| Black hole | Mass ($M_\odot$) | Angular diameter: $2\alpha_{\text{sh}}$ ($\mu$as) | Distance (kpc) |
| --- | --- | --- | --- |
| Sgr. A* | $4.3 \pm 0.013 \times 10^6$ (VLTI) | $48.7 \pm 7$ (EHT) | $8.277 \pm 0.033$ |
| M87* | $6.5 \pm 0.90 \times 10^9$ | $42 \pm 3$ | 16800 |

Sgr. A*: Sagittarius A*, M87*: Messier 87*, IAU: International Astronomical Union

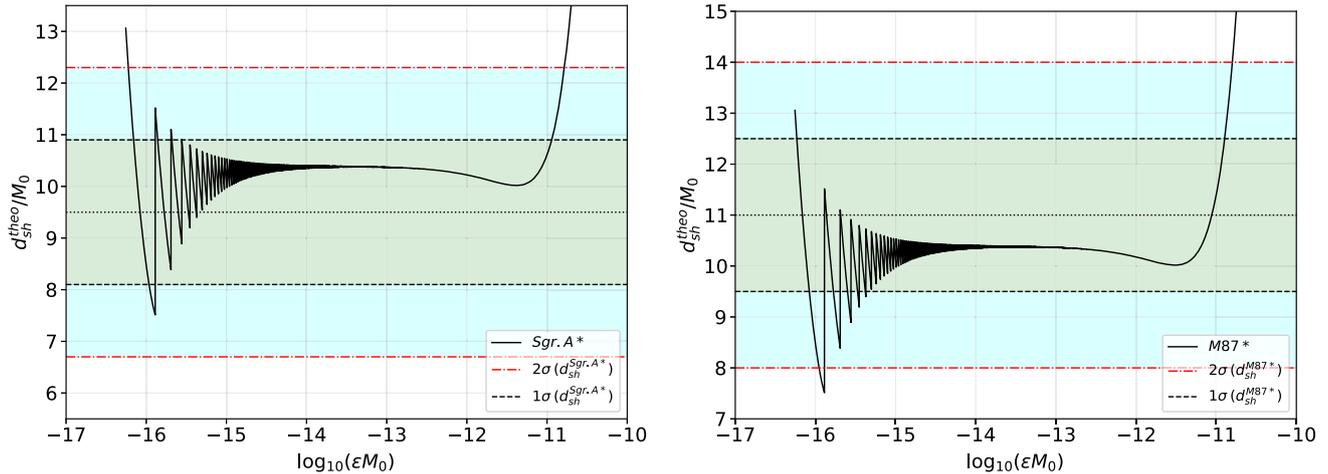

**Fig. 2** Constraints to $\epsilon$ based on EHT data. Left: For the $2\sigma$ level, there are two upper but no lower bounds. For $1\sigma$, there are several upper bounds but only one lower bound. Right: For the $2\sigma$ level, there are both upper and lower bounds. For $1\sigma$, there are two upper bounds but several lower bounds. Note that the horizontal black dotted line is the mean of the shadow diameter

**Table 2** Bounds of the dimensionless $\epsilon$ based on EHT results. The left table is for Sgr. A*, and the right one for M87*

| $\sigma$ level | Upper bound | Lower bound | $\sigma$ level | Upper bound | Lower bound |
| --- | --- | --- | --- | --- | --- |
| $2\sigma$ | $10^{-16.22}$ | – | $2\sigma$ | – | $10^{-15.89}$ |
|  | $10^{-10.78}$ | – |  | $10^{-10.80}$ | $10^{-15.95}$ |
| $1\sigma$ | $10^{-16.16}$ | $10^{-15.89}$ | $1\sigma$ | $10^{-16.23}$ | $10^{-15.37}$ |
|  | $10^{-10.94}$ | $10^{-15.97}$ |  | $10^{-10.89}$ | $10^{-16.08}$ |

With these data, one can find the diameter of the shadow size in units of the black hole mass with

$$d_{\text{sh}} = \frac{D\theta}{M_0}. \tag{38}$$

Thus, the diameter of the shadow image of M87* and Sgr. A* are $d_{\text{sh}}^{\text{M87*}} = (11 \pm 1.5)m$, and $d_{\text{sh}}^{\text{Sgr. A*}} = (9.5 \pm 1.4)m$ respectively. Meanwhile, the theoretical shadow diameter can be obtained via $d_{\text{sh}}^{\text{theo}} = 2R_{\text{sh}}$ where we used Eq. (37).

As seen in Fig. 2, the curve representing the diameter of the shadow radius decays in an oscillatory manner as $\epsilon$ increases. In Table 2, we indicated the values of the upper and lower bounds of $\epsilon$ for both uncertainty levels.

Due to the behavior of the curve, it can be concluded that there are multiple values of $\epsilon$ that give the same value of the shadow radius.

We also plotted how the shadow radius will behave relative to the position of a static observer from the black hole, as we use the bounds in the previous table for the parameter $\epsilon$. See Fig. 3. To avoid the scaling of $\epsilon$ and $\Lambda_0$, we use a log plot on the x-axis. We also compared the Schwarzschild and its modification as dS and AdS types to our scale-dependent model.

Indeed, the vertical dotted line represents the shadow radius in Fig. 2. Furthermore, there is no distinction between Schw-dS and AdS types at such a position unless one is located near the cosmological horizon. The Schw-AdS shadow merely shoots up indefinitely, while it can be seen that the Schw-dS shadow collapses to zero as one nears the cosmic horizon. We also note how such an observer so far away from the black hole can perceive the same shadow size as if he/she is near the horizon. For the scale-dependent BH under the parameter $\epsilon$, we see that the shadow is smaller for $\epsilon M_0 = 10^{-15.97}$, as compared to $\epsilon M_0 = 10^{-10.94}$. For Sgr. A*, we can see that either of these scale parameters mimics the AdS-type behavior, where a larger value produces the effect of the AdS type early on. Worth noting is how the shadow size varies at points very near our location. For low values of $\epsilon$, these fluctuations in the shadow size happen near the cosmic horizon. For M87*, the same conclusion can be formed, except for low values of $\epsilon$, where the AdS type scale dependent BH does not follow the behavior of the Schw-dS BH (see the dashed blue line).

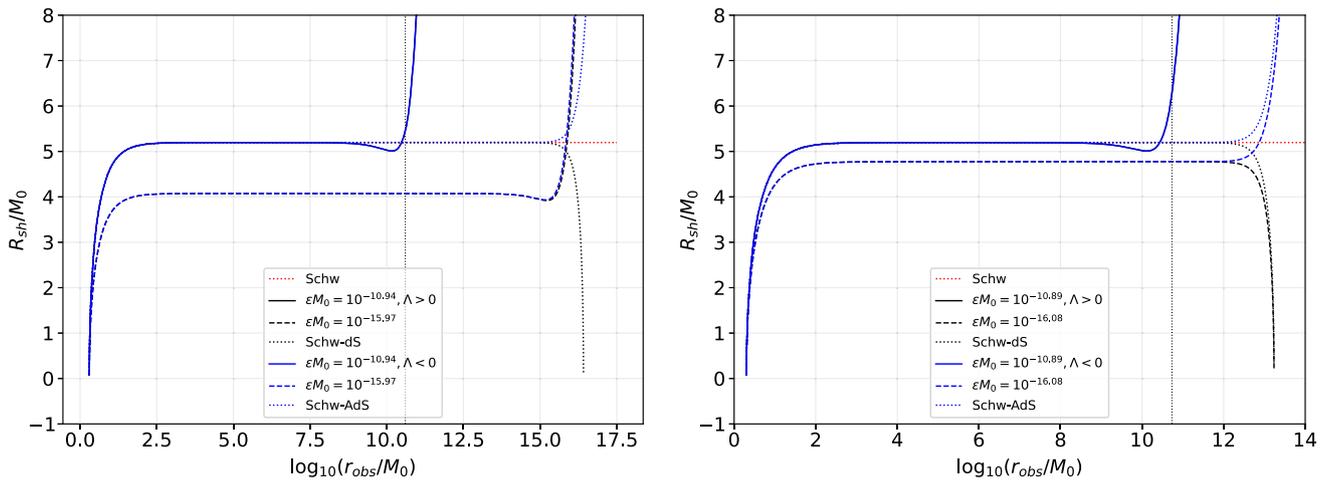

**Fig. 3** Here, $\Lambda_0 = 1.1 \times 10^{-52}$ m$^{-2}$. The vertical dotted line represents our location from the SMBH. The left and right figures are for Sgr. A* and M87*, respectively

## 4 Weak deflection of light of 4D scale-dependent Schwarzschild-AdS/dS black holes

In this section, we will probe the parameter $\epsilon$ and $\Lambda_0$ with weak deflection angle $\hat{\alpha}$. With this aim, we use the Gauss-Bonnet theorem, which states that [134, 135]

$$\iint_M K dS + \sum_{i=1}^N \int_{\partial M_a} \kappa_g d\ell + \sum_{i=1}^N \theta_i = 2\pi \chi(M), \quad (39)$$

where $K$ is the Gaussian curvature, $dS$ is the area measure, $\theta_i$, and $\kappa_g$ is the jump angles and geodesic curvature of $\partial M$, respectively, and $d\ell$ is the arc length measure. Its application to null geodesic at the equatorial plane implies that the Euler characteristics should be $\chi(M) = 1$. If the integral is evaluated over the infinite area surface bounded by the light ray, it was shown [136] that the above reduces to

$$\hat{\alpha} = \phi_{RS} + \Psi_R - \Psi_S = -\iint_{{}_R^\infty \square_S^\infty} K dS, \quad (40)$$

where $\hat{\alpha}$ is the weak deflection angle. In the above formula, $\phi_{RS} = \Psi_R - \Psi_S$ is the azimuthal separation angle between the source S and receiver R, $\Psi_R$ and $\Psi_S$ are the positional angles, and ${}_R^\infty \square_S^\infty$ is the integration domain. This study cannot use this formula since the spacetime considered here is non-asymptotically flat due to the cosmological constant term. Nevertheless, it was shown in [82] that if one uses the path in the photonsphere orbit instead of the path at infinity, the above can be recast in a form applicable for non-asymptotically flat spacetimes:

$$\hat{\alpha} = \iint_{{}_{r_{ps}}^R \square_{r_{ps}}^S} K dS + \phi_{RS}. \quad (41)$$

To determine $K$ and $dS$, consider that metric from an SSS spacetime

$$ds^2 = g_{\mu\nu} dx^\mu dx^\nu = -A(r)dt^2 + B(r)dr^2 + C(r)d\theta^2 + D(r)\sin^2\theta d\phi^2. \quad (42)$$

Due to spherical symmetry of the metric, it will suffice to analyze the deflection angle when $\theta = \pi/2$, thus, $D(r) = C(r)$. Since we are also interested in the deflection angle of massive particles, we need the Jacobi metric, which states that

$$dl^2 = g_{ij} dx^i dx^j = (E^2 - \mu^2 A(r))\left(\frac{B(r)}{A(r)}dr^2 + \frac{C(r)}{A(r)}d\phi^2\right), \quad (43)$$

where the energy per unit mass of the massive particle is

$$E = \frac{\mu}{\sqrt{1-v^2}}. \quad (44)$$

It is then useful to define another constant quantity in terms of the impact parameter $b$, which is the angular momentum per unit mass:

$$J = \frac{\mu v b}{\sqrt{1-v^2}}, \tag{45}$$

and with $E$ and $J$, we can define the impact parameter as

$$b = \frac{J}{vE}. \tag{46}$$

Using $ds^2 = g_{\mu\nu}dx^\mu dx^\nu = -1$, which is the line element for the time-like particles, the orbit equation can be derived as

$$F(u) \equiv \left(\frac{du}{d\phi}\right)^2 = \frac{C(u)^2 u^4}{A(u)B(u)}\left[\left(\frac{1}{vb}\right)^2 - A(u)\left(\frac{1}{J^2} + \frac{1}{C(u)}\right)\right], \tag{47}$$

which, in our case yields

$$F(u) = \frac{1}{v^2 b^2} + \left(\frac{1}{J^2} + u^2\right)\left[M_0(2u - 3\epsilon) + \frac{\epsilon}{u} + \frac{\Lambda_0}{3u^2} - 1\right]. \tag{48}$$

Here, $u = 1/r$ is usually done in celestial mechanics. Next, by an iterative method, the goal is to find $u$ as a function of $\phi$, which we find as

$$u(\phi) = \frac{\sin(\phi)}{b} + \frac{1+v^2\cos^2(\phi)}{b^2 v^2}M + \frac{b\Lambda_0}{6v^2} + \frac{\epsilon}{2v^2} + \frac{\epsilon M_0(v^2-1)}{2bv^4}. \tag{49}$$

The Gaussian curvature can be derived using

$$K = -\frac{1}{\sqrt{g}}\left[\frac{\partial}{\partial r}\left(\frac{\sqrt{g}}{g_{rr}}\Gamma^\phi_{r\phi}\right)\right], \tag{50}$$

since $\Gamma^\phi_{rr} = 0$ for Eq. (43). Furthermore, the determinant of Eq. (43) is

$$g = \frac{B(r)C(r)}{A(r)^2}(E^2 - \mu^2 A(r))^2. \tag{51}$$

With the analytical solution to $r_{co}$, it is easy to see that

$$\left[\int K\sqrt{g}\,dr\right]\bigg|_{r=r_{co}} = 0, \tag{52}$$

which yields

$$\int_{r_{co}}^{r(\phi)} K\sqrt{g}\,dr = -\frac{A(r)(E^2 - A(r))C' - E^2 C(r)A(r)'}{2A(r)(E^2 - A(r))\sqrt{B(r)C(r)}}\bigg|_{r=r(\phi)}, \tag{53}$$

where the prime denotes differentiation with respect to $r$. The weak deflection angle is then [82],

$$\hat{\alpha} = \int_{\phi_S}^{\phi_R}\left[-\frac{A(r)(E^2 - A(r))C' - E^2 C(r)A(r)'}{2A(r)(E^2 - A(r))\sqrt{B(r)C(r)}}\bigg|_{r=r(\phi)}\right]d\phi + \phi_{RS}. \tag{54}$$

Using Eqs. (49) in (53), we find

$$\left[\int K\sqrt{g}\,dr\right]\bigg|_{r=r_\phi} = -\phi_{RS} - \frac{(2E^2-1)M_0(\cos(\phi_R) - \cos(\phi_S))}{(E^2-1)b}$$
$$- \frac{\epsilon b \ln(\csc(\phi_R) - \csc(\phi_S) - (\cot(\phi_R) + \cot(\phi_S)))}{2(E^2-1)} + \frac{(1+E^2)b^2\Lambda_0(\cot(\phi_R) - \cot(\phi_S))}{6(E^2-1)}$$
$$- \frac{\epsilon M_0}{2v^2(E^2-1)}\left[(v^2+1)(\cot(\phi_R) - \cot(\phi_S)) + (2E^2-1)(v-1)(v+1)\phi_{RS}\right]. \tag{55}$$

We obtained the solution for $\phi$ as

$$\phi_S = \arcsin(bu) + \frac{M_0[v^2(b^2u^2-1)-1]}{bv^2\sqrt{1-b^2u^2}} - \frac{\epsilon b}{2v^2\sqrt{1-b^2u^2}} - \frac{b^2\Lambda_0}{6v^2\sqrt{1-b^2u^2}}$$
$$- \frac{\epsilon M_0\{bu[-u(bu+1)v^2 + bu + v^2 - 1] + v^2 - 1\}}{2v^4(1-b^2u^2)^{3/2}},$$

$$\phi_R = \pi - \arcsin(bu) - \frac{M_0[v^2(b^2u^2 - 1) - 1]}{bv^2\sqrt{1 - b^2u^2}} + \frac{\epsilon b}{2v^2\sqrt{1 - b^2u^2}} + \frac{b^2\Lambda_0}{6v^2\sqrt{1 - b^2u^2}}$$
$$+ \frac{\epsilon M_0\{bu[-bu(bu+1)v^2 + bu + v^2 - 1] + v^2 - 1\}}{2v^4(1 - b^2u^2)^{3/2}}. \tag{56}$$

With the above expression for $\phi$, we apply some basic trigonometric properties:

$$\cos(\pi - \phi_S) = -\cos(\phi_S), \quad \cot(\pi - \phi_S) = -\cot(\phi_S), \quad \sin(\pi - \phi_S) = \sin(\phi_S), \quad \csc(\pi - \phi_S) = \csc(\phi_S), \tag{57}$$

and we find the following:

$$\cos(\phi_S) = \sqrt{1 - b^2u^2} - \frac{M_0 u[v^2(b^2u^2 - 1) - 1]}{v^2\sqrt{(1 - b^2u^2)}} + \frac{b^2 u\epsilon}{2v^2\sqrt{1 - b^2u^2}} + \frac{b^3 u\Lambda_0}{6v^2\sqrt{1 - b^2u^2}}$$
$$- \frac{\epsilon M_0\{1 + bu - b^3u^3 + (bu - 1)(2bu - 1)[v(bu + 1)]^2\}}{2v^4(1 - b^2u^2)^{3/2}}, \tag{58}$$

$\cot(\phi_S)$ as

$$\cot(\phi_S) = \frac{\sqrt{1 - b^2u^2}}{bu} + \frac{M_0[v^2(-b^2u^2 + 1) + 1]}{b^3u^2v^2\sqrt{1 - b^2u^2}} + \frac{\epsilon}{2bu^2v^2\sqrt{1 - b^2u^2}} + \frac{\Lambda_0}{6u^2v^2\sqrt{1 - b^2u^2}}$$
$$- \frac{\epsilon M_0\{-b^3u^3 + 3b^2u^2 - (bu - 2)(bu - 1)[v(1 + bu)]^2 + bu - 2\}}{2b^3u^3v^4(1 - b^2u^2)^{3/2}}, \tag{59}$$

and $\ln(\csc(\phi_S) - \cot(\phi_S))$ as

$$\ln(\csc(\phi_S) - \cot(\phi_S)) = \ln\left(\frac{1}{bu}\right) + \ln(1 - \sqrt{1 - b^2u^2}) + \frac{[v^2(b^2u^2 - 1) - 1]M_0}{v^2b^2u\sqrt{1 - b^2u^2}} - \frac{\epsilon}{2uv^2(1 - b^2u^2)^2}$$
$$- \frac{b\Lambda_0}{6uv^2\sqrt{1 - b^2u^2}} - \frac{\epsilon M_0\{1 + v^2 + bu(v^2 - 1 - bu[bu(v^2 - 1) + v^2 + 2])\}}{2b^2u^2v^4(1 - b^2u^2)^{3/2}}. \tag{60}$$

Finally, using Eqs. (59)–(60) to Eq. (55), we derived the weak deflection angle for both time-like and null particles as

$$\hat{\alpha} \sim \frac{M_0(v^2 + 1)}{bv^2}\left(\sqrt{1 - b^2u_R^2} + \sqrt{1 - b^2u_S^2}\right) + \frac{b\Lambda_0(v^2 - 2)}{6v^2}\left[\frac{\sqrt{1 - b^2u_S^2}}{u_S} + \frac{\sqrt{1 - b^2u_R^2}}{u_R}\right]$$
$$+ \frac{\epsilon b(v^2 - 1)}{2v^2}\left[\operatorname{arctanh}\left(\sqrt{1 - b^2u_S^2}\right) + \operatorname{arctanh}\left(\sqrt{1 - b^2u_R^2}\right)\right]$$
$$- \frac{\epsilon M_0(v^4 - 1)}{2bv^4}\left\{\frac{\pi b(u_S + u_R)}{2} - b[u_S\arcsin(bu_S) + u_R\arcsin(bu_R)] + \sqrt{1 - b^2u_S^2} + \sqrt{1 - b^2u_R^2}\right\}, \tag{61}$$

which is general since it also admits finite distance of the source and the receiver from the black hole. Due to the existence of the third term in the above equation, $u_R$ and $u_S$ cannot be exactly zero. However, if we think of it as very close to zero, we can recast the above equation into its far approximation form, which is

$$\hat{\alpha} \sim \frac{2M_0(v^2 + 1)}{bv^2} + \frac{b\Lambda_0(v^2 - 2)}{6v^2}\left(\frac{1}{u_S} + \frac{1}{u_R}\right)$$
$$+ \frac{\epsilon b(v^2 - 1)}{2v^2}\left[\operatorname{arctanh}\left(\sqrt{1 - b^2u_S^2}\right) + \operatorname{arctanh}\left(\sqrt{1 - b^2u_R^2}\right)\right] - \frac{\epsilon M(v^4 - 1)}{bv^4}. \tag{62}$$

It is quite interesting that using the weak deflection angle, the scale parameter $\epsilon$ is only detectable for time-like particles since if $v = 1$,

$$\hat{\alpha} \sim \frac{4M_0}{b} - \frac{b\Lambda_0}{6}\left(\frac{1}{u_S} + \frac{1}{u_R}\right). \tag{63}$$

We plot the exact expression Eq. (61) as shown in Figs. 4 and 5.

Let us discuss first the case where the weak deflection angle by Sgr. A* as perceived by a static observer on Earth. When $v \to 1$, the effect of $\epsilon$ diminishes and what we are analyzing are close to the effects of the Schw-dS and AdS types. Note that the blue and black dotted lines overlap since the effects of $+\Lambda_0$ and $-\Lambda_0$ are nearly identical. Closer inspection will reveal that $-\Lambda_0$ produces a higher $\hat{\alpha}$ in this case, thus showing the power of weak deflection angle. The lowest point is where $\hat{\alpha}$ begins to manifest, and we can say that as $b/M_0$ increases, $\hat{\alpha}$ tends to go zero. As shown, the effect of $\epsilon$ is more evident for low impact parameters. Take note that we are using the deflection of massive particles, and we can see that it produces a higher value of $\alpha$. Again, it is very hard to distinguish the effect of the positive or negative cosmological constant since the lines overlap. Nonetheless, we can see that the higher *epsilon* is, the higher the weak deflection angle is. There are also values of $b/M_0$ where the two $\epsilon$ are indistinguishable from one another, showing where the curves intersect.

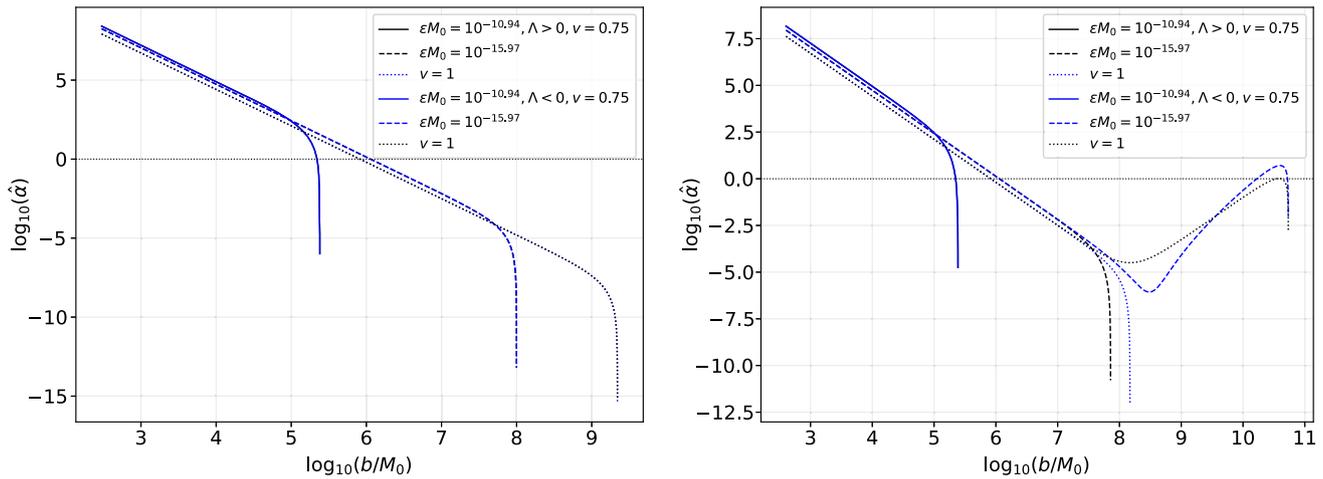

**Fig. 4** Behavior of the weak deflection angle in a log-log plot. The left figure is for Sgr. A*and the right one is for M87*

**Fig. 5** Behavior of the weak deflection angle as $u = 0.01b^{-1}$. Here, we used the parameters for M87*

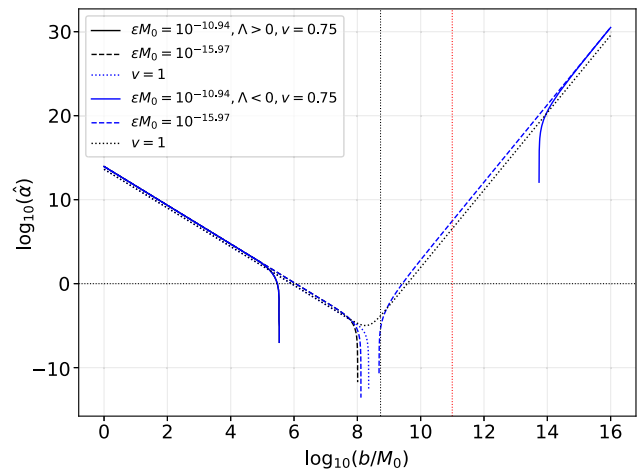

Next, let us discuss what happens to $\hat{\alpha}$ due to M87*. While some of the observations above apply to M87*, its mass and distance from us may give consequences to the weak deflection angle. Notice that at $\log_{10}(b/M_0) = 8$, we could see that it is possible to detect the difference between the Schw-dS and AdS types since the dotted curves decouple. It happens earlier for the dashed curves $\epsilon M_0 = 10^{-15.97}$, where the blue dashed curve extends to higher values of $b$. Therefore, we can see the detectability between the Schw-AdS and its scaled version. We note, however, that extreme precision of certain devices is needed since the weak deflection angle is small in this case.

Finally, let us discuss how the finite distance affects the weak deflection angle. In Fig. 5, we used $u = 0.01b^{-1}$ to show that there is a unique value of $u$ for every $b$. Thus, if $b$ is small, $u$ represents the inverse distance close to the black hole. Otherwise, it represents the distance that is so far from the black hole. With this assignment, each point in the plot is unique for every $u$ and $b$. Using our location from M87*, the corresponding impact parameter to be observed is $b = 5.18 \times 10^{21}$ m, which corresponds to 8.73 in the *x*-axis, represented by the vertical dotted line. Here, we can only differentiate between Schw-AdS and its scaled version. However, the weak deflection angle is rather small. Furthermore, the AdS type for Schw and scale-dependent BH increases $\hat{\alpha}$ indefinitely as one goes farther than the cosmological horizon. Finally, we note that the vertical red dotted line corresponds to the impact parameter if the source and the receiver are located at a distance comparable to the cosmological horizon. Here, we can only differentiate the Schw AdS and the AdS in scaled gravity. According to the plot, to observe the effect of the dS type in scaled gravity at high impact parameters, one must observe a black hole that goes far beyond the cosmological horizon, which is physically irrelevant.

## 5 Rigorous bounds of greybody factors

Since Hawking showed that black holes are not black [137], but because of emitting radiation, it is known as Hawking radiation and black holes look like grey objects [138–140]. An asymptotic observer can see this radiation differently than the original radiation near the horizon of the black hole due to the redshift factor, known as the greybody factor. There are various methods to obtain greybody factors [141–163], afterwards Visser and then Boonserm and Visser [164, 165] showed the elegant way to do it analytically and obtain the rigorous bounds on the greybody factors [166–176]. Here, we focus on bounds for the greybody factors of 4D scale-dependent Schwarzschild-AdS/dS black holes. To do so, first, we consider the Klein-Gordon equation for the massless scalar field as follows:

$$\frac{1}{\sqrt{-g}}\partial_\mu\left(\sqrt{-g}g^{\mu\nu}\partial_\nu\Phi\right) = 0. \tag{64}$$

Note that the solutions of the wave equation in spherical coordinates should be in the form of

$$\Phi(t, r, \Omega) = e^{i\omega t}\frac{\psi(r)}{r}Y_{\ell m}(\Omega), \tag{65}$$

with the spherical harmonics $Y_{\ell m}(\Omega)$. Then the Klein-Gordon equation reduces to this form:

$$\frac{\omega^2 r^2}{f(r)} + \frac{r}{\psi(r)}\frac{d}{dr}\left[r^2 f(r)\frac{d}{dr}\left(\frac{\psi(r)}{r}\right)\right] + \frac{1}{Y(\Omega)}\left[\frac{1}{\sin\theta}\frac{\partial}{\partial\theta}\left(\sin\theta\frac{\partial Y(\Omega)}{\partial\theta}\right)\right]$$
$$+ \frac{1}{\sin^2\theta}\frac{1}{Y(\Omega)}\frac{\partial^2 Y(\Omega)}{\partial\phi^2} = 0, \tag{66}$$

where the angular part should satisfy

$$\frac{1}{\sin\theta}\frac{\partial}{\partial\theta}\left(\sin\theta\frac{\partial Y(\Omega)}{\partial\theta}\right) + \frac{1}{\sin^2\theta}\frac{\partial^2 Y(\Omega)}{\partial\phi^2} = -\ell(\ell+1)Y(\Omega). \tag{67}$$

Note that the angular momentum quantum number, signified by $\ell$ and then the Klein-Gordon equation (Eq. (66)) is left with the radial part in the tortoise coordinate $r_*$:

$$\frac{d^2\psi(r)}{dr_*^2} + \left[\omega^2 - V(r)\right]\psi(r) = 0. \tag{68}$$

It is noted that $\frac{dr_*}{dr} = \frac{1}{f(r)}$. Then the effective potential $V(r)$ is obtained by

$$V(r) = \frac{\ell(\ell+1)f(r)}{r^2} + \frac{f(r)f'(r)}{r}. \tag{69}$$

Afterward, using the above effective potential, we study the lower rigorous bound for the greybody factor of the 4D scale-dependent Schwarzschild-AdS/dS black holes to probe the effect of $\epsilon$ on the bound. The formula to derive the rigorous bound of greybody factor is given as follows [164, 165]:

$$T_b \geq \text{sech}^2\left(\frac{1}{2\omega}\int_{-\infty}^{\infty}|V|\frac{dr}{f(r)}\right). \tag{70}$$

and the boundary of the above formula is modified for the cosmological constant [167] as follows:

$$T \geq T_b = \text{sech}^2\left(\frac{1}{2\omega}\int_{r_H}^{R_H}\frac{|V|}{f(r)}dr\right) = \text{sech}^2\left(\frac{A_\ell}{2\omega}\right), \tag{71}$$

with

$$A_\ell = \int_{r_H}^{R_H}\frac{|V|}{f(r)}dr = \int_{r_H}^{R_H}\left|\frac{\ell(\ell+1)}{r^2} + \frac{f'}{r}\right|dr. \tag{72}$$

We can calculate the bound numerically and plot it in Fig. 6 for $\ell = 0$ and Fig. 7 for $\ell = 1$. The graph shows that when the parameter of $\epsilon M_0$ increases, the bound of the greybody factor also decreases. It is found that the 4D scale-dependent Schwarzschild-AdS/dS black holes behave as good barriers and have a lower greybody factor than the Schwarzschild black hole.

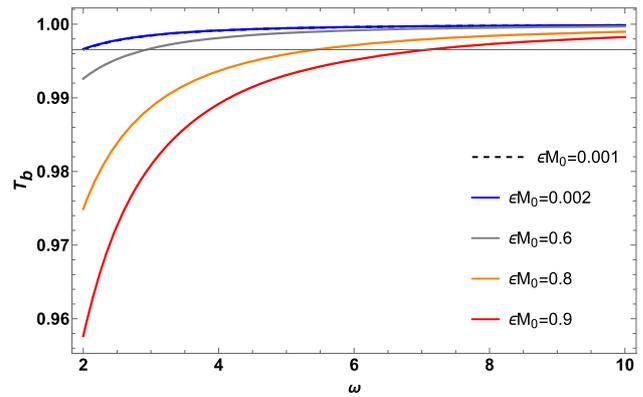

**Fig. 6** The greybody bound $T_b$ versus the $\omega$ for different values of $\epsilon$, with $M_0 = 1$, $\ell = 0$, $s = 0$, and $\Lambda_0 = 0.0001$

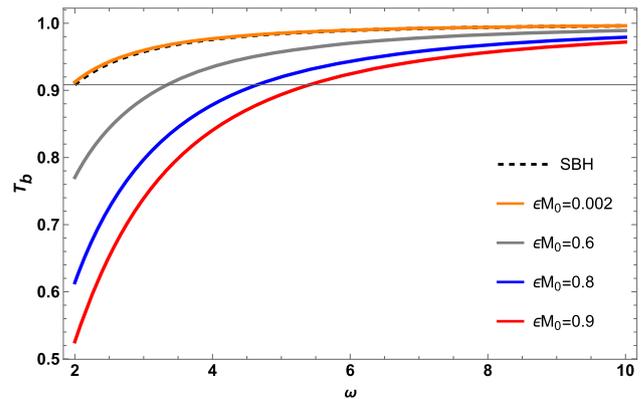

**Fig. 7** The greybody bound $T_b$ versus the $\omega$ for different values of $\epsilon$, with $M_0 = 1$, $\ell = 1$, $s = 0$, and $\Lambda_0 = 0.0001$

## 6 Conclusion

In this paper, we have investigated the shadow and weak deflecting angle of 4D scale-dependent Schwarzschild-AdS/dS black holes for the first time. Also, its greybody bound was computed too. We first find constraints for $\epsilon$ using the EHT data. These values are summarized in Table 2. We then applied such bounds to analyze the shadow cast behavior as perceived by a static observer, whose position $r_{obs}$ can be found near the black hole or to a limit near the cosmological horizon. The Schw dS and AdS types manifest at points near such limits. However, we have seen that $\epsilon$'s effect is to disrupt such behavior. For instance, as shown in Fig. 3, the effect of high $\epsilon$ is to make the AdS type behavior become a dS type relative to our location from Sgr. A* or M87*. For M87*, we also observe a decoupling behavior between the dS and AdS Schw in 4D scaled gravity.

Next, we have implemented the Gauss-Bonnet theorem, modified to allow calculation of the weak deflection angle for non-asymptotically flat spacetime. We derived an expression that accommodates finite distance, massive, and null particles. We have found that as $v \to 1$, the weak deflection by a BH in 4D scaled gravity approaches the Schw dS and AdS type. That is, the effect of the scale parameter $\epsilon$ vanishes. However, observing the deflection angle of relativistic particles, we have seen that such deviations from the standard ones only manifest at low values of impact parameters for Sgr. A*. For M87*, the effect of AdS scaled gravity can also manifest at high impact parameters at the expense of low values of $\hat{\alpha}$. Consequently, ultra-sensitive devices are needed to detect the effect of scaled gravity at high impact parameters. Finally, we also considered an alternative analysis where it would be possible to determine $\hat{\alpha}$ if both the light source and the receiver are near the cosmological horizon. Our results revealed that the resulting value $\hat{\alpha}$ is even higher than those at low impact parameters. The effects of $\epsilon$ can still be realized in this domain, except for the scaled dS type model. See Fig. 5.

In Einstein's theory of relativity, it is known that greybody factor is related to a black hole's quantum nature, so a high value of the greybody factor shows a high probability that Hawking radiation can propagate to infinity. In this way, one can extract quantum information from the greybody factor to gain insight into the theory of quantum gravity. In this paper, last we studied the rigours bound of the greybody factor numerically and plot it in Fig. 6 for $\ell = 0$ and Fig. 7 for $\ell = 1$. We show in the graph that when the parameter of $\epsilon M_0$ increases, the bound of the greybody factor also decreases. The numerical analysis demonstrates that the effect of parameter $\epsilon M_0$ estimation is significant when the deflection angle, the shadow cast, and the rigorous bound of the greybody are calculated. Future observations and further analysis will test the shadow cast's stability, shape, and depth more accurately and give us hints about the black holes in light of scale-dependent gravity.

**Acknowledgements**   A.R. is funded by the María Zambrano contract ZAMBRANO 21-25 (Spain). A. Ö. and R. P. would like to acknowledge networking support by the COST Action CA18108 - Quantum gravity phenomenology in the multi-messenger approach (QG-MM).

**Funding**   Open Access funding provided thanks to the CRUE-CSIC agreement with Springer Nature.

**Data Availability Statement**   We do not have any additional data to present.